\documentclass[aps,prd, twocolumn ,
showkeys,showpacs,amssymb,
amsfonts,epsf,
preprintnumbers,nofootinbib,superscriptaddress]{revtex4}

\usepackage[dvips]{graphicx}
\usepackage{bm,latexsym,amsmath,amssymb,amsfonts}
\usepackage{color}

\newcommand\beq{\begin{eqnarray}}
\newcommand\eeq{\end{eqnarray}}

\begin{document}

\title{Field localization on brane intersections in anti-de Sitter spacetime}
\author{Antonino Flachi}
\affiliation{Yukawa Institute for Theoretical Physics, Kyoto University,
Kitashirakawa Oiwake-cho, Kyoto 606-8502, Japan}
\author{Masato~Minamitsuji}
\affiliation{Arnold-Sommerfeld-Center for Theoretical Physics, Department f\"{u}r Physik,
Ludwig-Maximilians-Universit\"{a}t, Theresienstr. 37, D-80333, Munich, Germany}
\affiliation{Center for Quantum Spacetime,
Sogang University, Shinsu-dong 1, Mapo-gu, 121-742 Seoul, South Korea}
\email{minamituzi_at_sogang.ac.kr}

\begin{abstract}
We discuss the localization of scalar, fermion, and gauge field zero modes on a $3-$brane that resides at the intersection of two $4-$branes in six-dimensional anti-de Sitter space. This set-up has been introduced in the context of brane world models and, higher-dimensional versions of it, in string theory.
In both six- and ten-dimensional cases, it has been shown that four-dimensional gravity can be reproduced at the intersection, due to the existence of a massless, localized graviton zero-mode. However, realistic scenarios require also the standard model to be localized on the $3-$brane.
In this paper, we discuss under which conditions a higher-dimensional field theory, propagating on the above geometry, can have a zero-mode sector localized at the intersection and find that zero modes can be localized  only if masses and couplings to the background curvature satisfy certain relations.
We also consider the case when other 4-branes cut the bulk at some distance from the intersection and argue that, in the probe brane approximation, there is no significant effect on the localization properties at the $3-$brane.
The case of bulk fermions is particularly interesting, since the properties of the geometry allow localization of chiral modes independently.
\end{abstract}
\pacs{04.50.+h, 98.80.Cq}
\preprint{YITP-09-07}
\preprint{CQUeST-2009-0263}
\maketitle
\date{today}


\section{Introduction}

Branes appear in a variety of physical contexts, and, during the past years, have become one of the basic building blocks of higher-dimensional cosmological and particle physics models, due to their property to confine to three dimensions both gravitational and nongravitational fields via a  mechanism of localization \cite{ArkaniHamed:1998rs,Randall:1999ee}.

A $p-$brane is a $p-$dimensional extended object that sweeps out a $(p+1)-$dimensional
worldvolume, as it evolves in time, in a $d-$dimensional bulk space, essentially playing the role of {\it boundary conditions} for the fields propagating in the bulk. In the compactification scheme, a four-dimensional effective field theory is the result of a spectral reduction of the higher-dimensional field into a
zero-mode plus massive excitations. However, when branes are present, a radical difference arises due to the fact that branes effectively separate the bulk space
into two (or more) sectors, requiring specific conditions to be imposed on the fields at the junctions.
When the bulk geometry is appropriately chosen, the effect of the boundary conditions is to localize the zero-mode and eventually massive excitations on the branes.
By the same procedure as in the compactification scheme, the zero-mode of an appropriately chosen
higher-dimensional field theory may be {\it matched} to the standard model spectrum.
Depending on the details of the model, essentially tuning some scales of the bulk space,
the masses of the higher-dimensional excitations can be made of order of some TeV, thus producing modifications to standard four-dimensional physics with possible phenomenological or cosmological signatures.
In some cases (see Ref.~\cite{Randall:1999vf}), the presence of branes allows extra dimensions to be taken infinitely large without any confliction with observation. In this case, the spectrum consists of a continuum of modes.

Choosing bulk geometry, number of branes, and arranging them in the bulk suffers, in principle, from a degree of arbitrariness, and different features may produce different low-energy effective theories and lead to different cosmological or phenomenological predictions.
For this reason, a case by case analysis of the low-energy structure of the model seems required.
However, the basic properties of a brane universe are rather generic.
For instance, as it was shown in \cite{addk}, the mechanism of localization due to the warping of the anti-de Sitter (AdS) bulk space {\it only} requires the presence of codimension one branes. Thus, extending the model to more than five dimensions, localization can be realized by placing a $3-$brane at the intersection of a number of co-dimension one $p-$branes. This remains true also when all the branes have tension, as it was discussed in Ref.~\cite{origamido}, where it was shown how to
construct
a brane universe by gluing together patches of six-dimensional $AdS$ space \cite{origamido}.
In the  model of Ref.~\cite{origamido} the observed universe is a $3-$brane located at the intersection of two $4-$branes, all embedded in a six-dimensional $AdS$ bulk, leading to a pyramidlike structure.
Minkowski, de Sitter and anti-de Sitter geometries can be realized on the 3-brane, and ordinary four-dimensional gravity can be reproduced. Also, as in scenarios with large or warped extra-dimensions, the effective scale of gravity can be reduced to a few TeV resolving in a geometrical way the hierarchy problem.
The cosmology in six-dimensional intersecting brane models has also received some recent attention (see, for example, \cite{cor1}).

Another recent application of intersecting brane set-ups has been considered in Refs.~\cite{Karch:2005yz,fizran}, that argue that in ten-dimensional string theory on an AdS background, $3-$ and $7-$branes could come to dominate the cosmological evolution of the bulk space.
\footnote{
Some related arguments claiming that in the context of string theory {\it only} D3-branes are 
cosmologically favored have been put forward for instance 
in Ref.~\cite{md,dks}.
Refs.~\cite{md} discussed the possibility of
the decay of initial space-filling D$9$-$\bar D9$ brane
pairs, via tachyon dynamics, which may lead to phase transitions, dynamics
and creation of the D-branes,
and Ref. \cite{dks} considered
intersections leading to reconnection and unwinding of intersecting D$p$-branes ($p > 3$), 
although in these works no mechanism to localize the gravity on one of the 3-branes has been discussed.}
If so, four-dimensional gravity can
be reproduced on the triple intersections of $7-$branes, offering a possible explanation for the observed three dimensionality of space. The localization of gravity in this model has been discussed in Ref.\,\cite{fizran}, showing that the model has analogous properties to those of the equivalent six-dimensional model of \cite{origamido}.

Intersecting brane models are also very popular in string theory with flat compactified extra dimensions,
and they are usually constructed from $D6-$branes intersecting on a
four-dimensional
manifold and wrapping different cycles in the transverse compact space.
Their popularity arises because
chiral fermions, localized at the intersection, can be easily obtained.
Intersecting brane models have also various other bonuses like replication of quark-lepton generation.
They may lead to hierarchical structures for quark and lepton masses easily, the hierarchy problem can
also be addressed in the standard way by taking the volume of the extra dimensions large enough.
The fact that the masses of the excited modes are proportional to the angles of the intersecting branes
and may be just above the weak scale, make such models sensitive to collider experiments, and motivated
much work devoted to study their physics (see, for example, \cite{douglas,ibanez,friedel}).

In this paper, for simplicity, we will focus on a model that consists of a $6-$dimensional bulk with two $4-$branes intersecting along a $3-$brane. Along with this case, we also discuss the effect of a extra $4-$branes, playing the role of cut-off in the extra dimensions. These extra branes require specific boundary conditions to be obeyed at their surface by the fields propagating in the bulk, and, as we will argue, adding cut-off branes, and thus having a finite volume bulk, does not change the behavior of the localized zero modes substantially.
The geometry considered in this paper will be introduced, mainly following Ref.~\cite{origamido}, in the next section.

Here, we are interested in studying what are the conditions of localization at the intersection for the zero-mode of a higher dimensional field.
Let us first briefly summarize the situation arising when branes do not intersect.
The corresponding model to the scenario of Ref.~\cite{origamido} is the Randall-Sundrum model, with an infinitely extended five-dimensional AdS bulk,
and one brane. In this case, scalar fields with arbitrary boundary mass terms, can be localized on a positive tension brane, and have continuous spectrum with no mass gap. Half-integer, $1/2$ and $3/2$, spin fields can instead be localized on a negative tension brane. Their localization to a positive tension brane can be realized at the price of introducing some specific boundary mass terms. Massless, spin $1$ localized zero modes do not arise from a higher-dimensional vector field, but they may arise from higher-dimensional antisymmetric forms. Adding extra parallel branes, does not change the localization issue, but discretize the spectrum. This has been at the center of much attention and has been suggested of a mechanism to explain, for example, the fermion mass hierarchy, the smallness of neutrino masses, or a way to give to the mechanism of supersymmetry breaking a geometrical origin. We refer the reader to the literature, for example
Refs.
\cite{Bajc:1999mh,Gherghetta:2000qt,ran,rub,mirab,neub,kogan,ring,sing,jacky,pomarol}
where the issue of localization in the Randall-Sundrum model with one or two branes
has been studied, and to Ref.~\cite{kogan} where a multi brane set-up has been considered.
(See \cite{liu} for an example of localization on thick branes.)
It seems interesting to extend the above results to the case of brane intersections in AdS, and in Sec. III we will discuss this. We will consider a theory with general masses and coupling to the curvature and illustrate that a localized zero-mode sector is possible only if masses and couplings obey certain relations. We will also show that in the case of fermions, different chiral modes can be localized separately, thus the geometrical mechanism of generating mass hierarchies in the Randall-Sundrum model, can be extended to this case.

Sec. IV will be devoted to briefly consider the effects of the Kaluza-Klein excitation and in particular the mass spectrum, which, as shown in \cite{origamido} for the case of the graviton, is continuous and gapless.
The couplings of the excited modes with the localized zero modes are, however, negligible thus viable phenomenologies can in principle be constructed. This can be seen from the profile of the potential, which forms a barrier around the intersection protecting the branes from potentially dangerous modes.

\section{Intersecting Branes in AdS}

\label{secII}

In this section we will present the prototype background geometry, briefly summarizing the results of \cite{origamido}.
The system consists of a six-dimensional
bulk, two 4-branes which are intersecting in the bulk,
and a 3-brane residing at the intersection.
The bulk action is given by
\beq
S_{bulk} =- \int d^6X \sqrt{-g} \Big({M^4_6\over 2}R +\Lambda\Big)~.
\eeq
We will set $M_6$ to be unity for the moment.
Two tensional 4-branes
intersect along a
3-brane.
Then, the brane action is given by
\beq
S_{brane} = \sum_{i=1}^2 \int_{\Sigma_i} d^5x \sqrt{-q_i}{\cal L}_i
   +\int_{\Sigma_0} d^4x \sqrt{-q_0}{\cal L}_0,
\eeq
with $q_{(1,2)}$ being the induced metric on the 4-branes, and $q_0$ on the $3-$brane.
As the 4-brane and 3-brane matter, we will consider only tensions
\beq
{\cal L}_i=-\sigma_i,\quad
{\cal L}_0=-\sigma_0.
\eeq

The bulk theory contains an anti-de Sitter solution,
whose metric can be written in the form
\beq
ds^2 = A(t,z_1,z_2)^2\left( -\delta_{mn} dz^mdz^n +\eta_{\mu\nu} dx^{\mu}dx^{\nu}\right)~,\label{metric}
\eeq
where $\eta_{\mu\nu}={\rm diag}(+1,-1,-1,-1)$ and the warp factor is
$$
A(t, z_1,z_2) = {1\over  Ht+k_1z_1 +k_2z_2+C_0}~.
$$
The constants $k_1$ and $k_2$ parametrize the bulk curvature
along two orthogonal directions, $z_1$ and $z_2$.
The constant $C_0>0$ is introduced in order for the 3-brane
not to be located at the position of AdS horizon.
In the above metric, $\eta_{\mu\nu}$ and $x^\mu$ represent the standard four-dimensional flat space metric and coordinates.

Let us construct the exact bulk metric which is locally AdS,
and contains two (intersecting) 4-branes and a 3-brane at
their intersection.
We assume that
two boundary $4-$branes exist along the lines $z^2=\tan \alpha_1 z^1$
and $z^2=-\tan \alpha_2 z^1$ on $(z^1,z^2)$-plane ($z^1>0$),
and
intersect at an angle and thus their relative `inclination' is characterized by two normal vectors
$
{\bf n}^{(1)}=(\sin\alpha_1,-\cos\alpha_1)$
and
$
{\bf n}^{(2)}=(\sin\alpha_2,\cos\alpha_2).
$
The intersection point is assumed
to be at $z^1=z^2=0$.
A set of two vectors parallel to the $4-$branes (orthogonal to the normal vectors),
${\bf l}_{(i)}\cdot  {\bf n}^{(j)}
=\delta_i{}^j~
$
can be introduced
${\bf l}_{(1)}
=\big(\cos\alpha_2/ \sin\alpha,-\sin\alpha_2/ \sin\alpha\big)~$ and
${\bf l}_{(2)}
=\big(\cos\alpha_1/ \sin\alpha,\sin\alpha_1/\sin\alpha\big)~,
$
where ${\bf l}_{(1,2)}$ are parallel to the branes $(2,1)$, respectively
(See e.g., Fig. 1 of Ref. \cite{origamido}).
Note that $\alpha \equiv \alpha_1+\alpha_2$.
A more convenient coordinate system $\tilde z\equiv(u,v)$ can be defined as
${\tilde z}^k \equiv {\bf n}_{(i)}\cdot {\bf z}~,$ where
\begin{eqnarray}
u= \sin \alpha_1 z^1-\cos \alpha_{1}z^2,\quad
v=\sin \alpha_2 z^1+\cos \alpha_2 z^2. \nonumber
\end{eqnarray}
The relation between the two coordinate systems can be summarized as
$
{\bf z}={\tilde z}^{k}\cdot {\bf l}_{(k)},\,\,
\partial z^h/\partial {\tilde z}^k=\ell^h{}_{(k)}~
$.
And it is easy to see that, in the $\tilde z$ coordinates, the bulk metric takes the form
\begin{eqnarray}
ds^2=A(t,u,v)^2
\Big(
-\gamma_{mn}d\tilde z^m d\tilde z^n
+\eta_{\mu\nu}dx^{\mu}dx^{\nu}
\Big),
\label{tcrds}
\end{eqnarray}
where $A(t,u,v)=1/(Ht+C_1u+C_2v+C_0)~$
and we define
\beq
C_1:=\frac{k_1\cos\alpha_2-k_2\sin\alpha_2}{\sin\alpha}~,\\
C_2:=\frac{k_1\cos\alpha_1+k_2\sin\alpha_1}{\sin\alpha}~.
\eeq
The components of the metric tensor in the transverse directions are
$
\gamma_{11}=\gamma_{22}=1/ \sin^2\alpha
$
and
$\gamma_{12}=\gamma_{21}=
\cos\alpha / \sin^2\alpha~$.
In the $\tilde z$ coordinates the $4-$branes are located at $u=0$ and $v=0$.
In order for the metric Eq. (\ref{tcrds}) to represent
a region of the bulk spacetime surrounded by the 4- and 3-branes,
we pick up a patch of AdS, $u>0$ and $v>0$.

To realize the intersecting brane system, we are going to
orbifold the bulk region.
The bulk spacetime can be constructed
by gluing four copies of the above patch of AdS spacetime
with $Z_2$-symmetry across each 4-brane.
Then, the coordinates $\tilde z$ can cover the whole bulk,
where the domain of both $u$ and $v$ coordinates is extended to be
$-\infty <u <\infty$ and $-\infty<v<\infty$.
The global bulk metric is given by Eq. (\ref{tcrds})
with redefinitions such that
$$
A(t,u,v):={1\over Ht+C_1|u|+C_2|v|+C_0}~
$$
and
$$
\gamma_{11}:=\gamma_{22}={1\over \sin^2\alpha},
$$
$$
\gamma_{12}:=\gamma_{21}=s(u) s(v)
{\cos\alpha \over \sin^2\alpha~}.
$$
We label each patch of bulk space as
\beq
\mbox{(I)}&&u>0\quad v>0, \nonumber \\
\mbox{(II)}&&u<0\quad v>0, \nonumber \\
\mbox{(III)}&&u<0 \quad  v<0,\nonumber \\
\mbox{(IV)}&&u>0 \quad v<0.
\eeq
The 4-branes are along the edges and the 3-brane is at the apex in the {\it pyramidal} bulk.

Requiring that the above spacetime is a solution of Einstein equations
gives a relation between the constants $k_1$ and $k_2$, the bulk cosmological constant and the Hubble constant
$$
\Lambda = 10(H^2-k_1^2-k_2^2)~.
$$
Brane tensions are determined through the boundary parts
of the Einstein equation, leading to the following relations\footnote{We thank Kazuya Koyama for pointing out an error on the 3-brane tension.}
\beq
\kappa_6^2 \sigma_1&=&8\big(C_1-C_2\cos\alpha\big)~,\nonumber\\
 \kappa_6^2
 \sigma_2&=&8\big(C_2-C_1\cos\alpha\big)~,\\
\kappa_6^2\sigma_0&=&4\Big(\frac{\pi}{2}-\alpha\Big)~.
\nonumber
\label{ten}
\eeq
In the rest of the paper, we focus on Minkowski 3-brane solutions and therefore set $H=0$, thus we define
$$
A(u,v):= A(0,u,v)~.
$$
Note that a de Sitter 3-brane with $H\neq 0$
can be obtained via boosting
one of the 4-branes in a particular direction
of the extra dimensions.

It is natural to assume the presence of additional $3-$ and $4-$branes in the above set-up.
Note that
any configuration of multiple brane intersections
must be a solution of Einstein equations
plus a set of consistency conditions,
which is generalization of the results obtained in \cite{gkl}.
We will briefly consider, along with the above geometrical construction, the effect of extra $4-$dimensional probe-branes, which cut the bulk space at some distance $\ell$ from the intersection.
As we will see later,
for these cut-off branes, the effect on the zero-mode localization is not substantial.


\section{Bulk Fields and Zero-Mode Localization.}

This section will be devoted to consider a $U(1)$ gauge field, $A_{\mu}$, a scalar field, $\phi$, and a Dirac fermion, $\Psi$, propagating in the above intersecting brane geometry in AdS. The starting action is the standard one
\begin{widetext}
\beq
S=
\int d^6X \sqrt{-g}
&\Big[&-{1\over 4}F_{MN}^2
+
\left((M_A^2-\chi R )g^{MN}- \tau R^{MN} \right)A_M A_N
+{1\over 2}\left(
(\partial \phi)^2
-\left(M_s^2
-\xi R\right) \phi^2
\right)\nonumber\\
&+&i\bar{\Psi}\left(\underline{\gamma}^MD_M + M_{f}\right)\Psi
\Big].
\label{action}
\eeq
In the above action, $F_{MN}=\partial_M A_N-\partial_N A_M$, the coefficients $M_A$, $M_s$ are the masses of the gauge and  scalar fields, respectively. Since we are in six dimensions, the Dirac spinor is eight dimensional and therefore $M_f$ is a $8\times 8$ matrix, whose form is restricted by the symmetries of the action. The constants $\chi$, $\tau$ and $\xi$ represent the coupling to the curvature. The covariant derivative is $D_M=\partial_M+\Gamma_M$, with $\Gamma_M$ being the spin connection. In the present paper, we ignore the effects of the back reaction.
The remainder of the section will be devoted to discuss the possibility to localize the zero-mode sector of the above higher-dimensional field theory on the $3-$brane.

\subsection{Scalars}
\label{secIII}

The field equations for the scalar field can be written in the usual Klein-Gordon form, that in the case of the intersecting brane geometry (\ref{metric}), take the form
\beq
\Big[
-
\Box
 +{\partial^2\over \partial u^2}+{\partial^2\over \partial v^2}
-{2 \cos \alpha \over s(u)s(v)}{\partial^2\over \partial u \partial v}
-2\cos\alpha \delta (u)s(v)\partial_v
-2\cos\alpha \delta(v) s(u)\partial_u
-V(u,v)
\Big]\Phi=0~,\label{op}
\nonumber
\eeq
where, for convenience, we have rescaled the field: $$\Phi=\big(C_1|u|+C_2|v|+C_0\big)^{-2}\phi.$$ In the above expression, the D'Alembertian is the standard one in four-dimensional Minkowski space-time, and the symbol $s(x)$ is used to indicate the sign-function. The potential term, $V(u,v)$, consists of a bulk part plus two $\delta-$function contributions coming from branes
\beq
V(u,v)&:=&{\tilde M^2} A^{2}(u,v)
- h_1(u,v) \delta(u)
-  h_2(u,v) \delta(v)~
+8\xi 
\Big(\frac{\pi}{2}-\alpha \Big)
\delta(u)\delta(v)
,\label{potential}
\eeq
with $\tilde M^2:= M_s^2+30(1/5-\xi)\big(k_1^2+k_2^2\big)$ and
\beq
h_1(u,v) := 4\big(C_1-C_2\cos\alpha\big)(1-5\xi)A(u,v)~,~~
h_2(u,v):= 4\big(C_2-C_1\cos\alpha\big)(1-5\xi)A(u,v)~.
\eeq
\begin{figure}
\begin{minipage}[t]{.55\textwidth}
\label{fig1}
   \begin{center}
    \includegraphics[scale=.75]{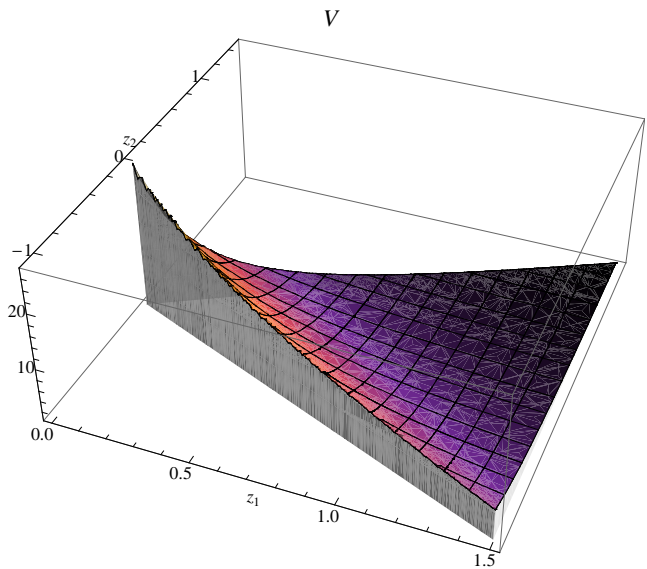}
\vspace{.0cm}
        \caption{
Scalar field potential $V$ defined in Eq. (\ref{potential}), shown in one of the four patches of $AdS_6$.
The figure illustrates the presence of a potential barrier around the 3-brane.
The 3-brane is located at the origin $z_1=z_2=0$
and 4-branes correspond to the boundaries (sides in the plot).
Here, we set $\alpha_1=\pi/5$ and $\alpha_2=\pi/4$.
Note that
the behavior of the potential for the vector field $V_A$
defined in Eq. (\ref{potential_A})
is very similar to the scalar one.
}
   \end{center}
 \end{minipage}
\end{figure}
\hspace{0.4cm}
\begin{figure}
\begin{minipage}[t]{.40\textwidth}
\label{fig2}
   \begin{center}
    \includegraphics[scale=.55]{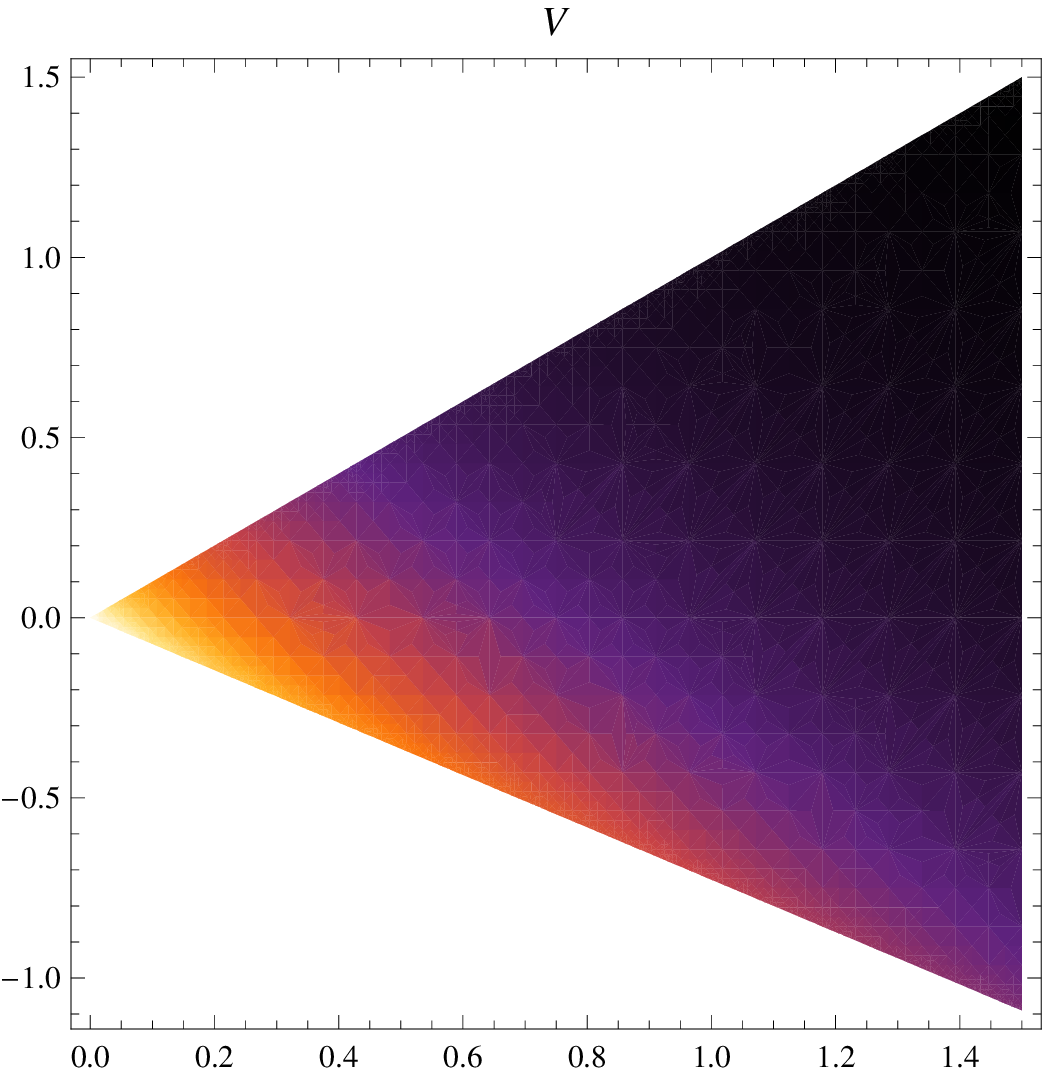}
\vspace{+.0cm}
        \caption{
A contrast of the potential is shown
for the same parameters as in Fig. 1.
}
   \end{center}
   \end{minipage}
\hspace{0.5cm}
\begin{minipage}[t]{.40\textwidth}
\label{fig3}
   \begin{center}
    \includegraphics[scale=.75]{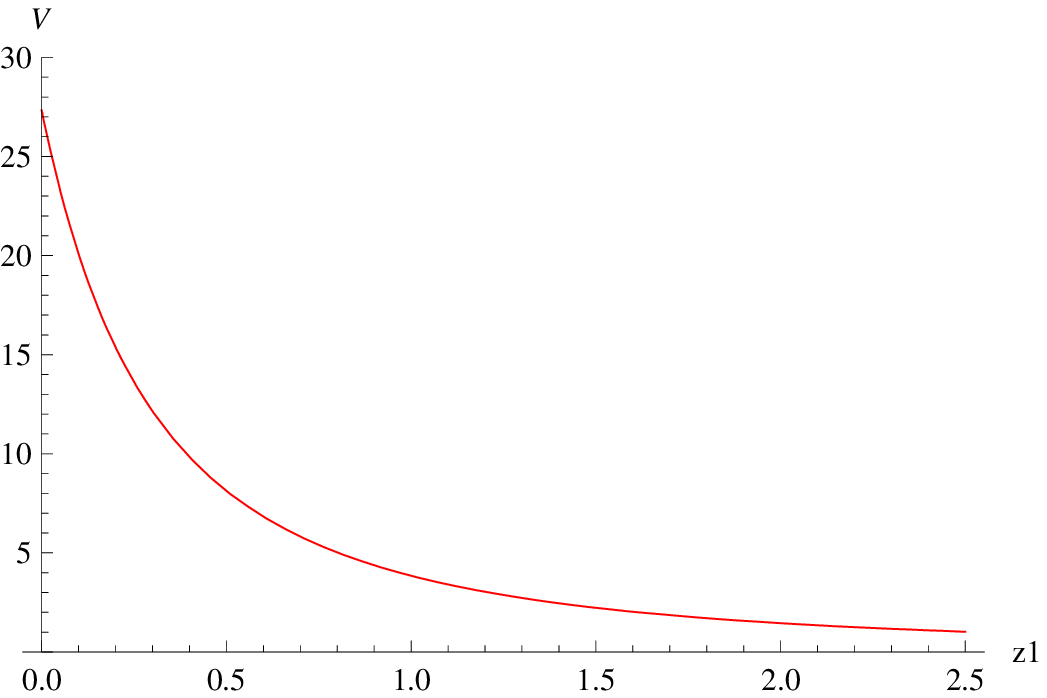}
\vspace{+.0cm}
        \caption{
The behavior of the potential along the $z_1$-axis
is shown for the same parameters as in Fig. 1.
}
   \end{center}
 \end{minipage}
\end{figure}
The four-dimensional effective theory can be obtained in the standard way by integrating out the extra dimensions. First, we decompose the higher-dimensional field, separating out the massless zero-mode from the Kaluza-Klein (KK) excitations
\begin{eqnarray}
\Phi= f_{0}(u,v) \varphi_{0}(x^\mu)
+\int d\lambda f_{\lambda}(u,v) \varphi_{\lambda}(x^\mu)\,.
\end{eqnarray}
Then, we assume that each mode $\varphi_{\lambda}$ satisfies a Klein-Gordon equation in four dimensions,
\beq
\Box \varphi_{\lambda} = - m_{\lambda}^2 \varphi_{\lambda} ~,
\label{4Det}
\eeq
with $\lambda$ being a multi-index. Requiring the modes to be normalizable,
\beq
\int dudv\frac{1}{\sin\alpha} f_{0}(u,v)f_{0}(u,v) = 1,~~~
\int dudv\frac{1}{\sin\alpha}  f_{\lambda}(u,v)f_{\lambda'}(u,v) =
\delta(\lambda -\lambda'),
\label{norm}
\eeq
a simple calculation shows that the six-dimensional action reduces to
\beq
 S&=&
-
\frac{1}{2}
\int d^4 x\varphi_{0}
    \Box \varphi_{0}
-
\frac{1}{2}\int d\lambda
\int d^4 x\varphi_{\lambda}
    \Big(\Box
+
m_{\lambda}^2
\Big)
\varphi_{\lambda}~,
\label{eftaction}
\eeq
where $m_\lambda$ are the masses of the KK excitations. The functions $f_{\lambda}(u,v)$ satisfy
\beq
\Big[
{\partial^2\over \partial u^2}+{\partial^2\over \partial v^2}
-{2 \cos \alpha \over s(u)s(v)}{\partial^2\over \partial u \partial v}
-2\cos\alpha \delta (u)s(v)\partial_v
-2\cos\alpha \delta(v) s(u)\partial_u
-V(u,v)+m_{\lambda}^2
\Big]f_{\lambda}=0~.
\label{radeqt}
\eeq
When supplemented with appropriate boundary conditions, solving the above equation allows to determine the masses of the KK excitations. The
four- dimensional effective theory, on the $3-$brane, is then completely described by the action (\ref{eftaction}).
The boundary conditions can be easily derived by integrating
the equation of motion for each bulk mode Eq. (\ref{radeqt})
across the $4-$branes.
The boundary conditions between regions $(I)$ and $(II)$
and between regions $(III)$ and $(IV)$ are given by
\beq
 \frac{\partial f_{\lambda}}{\partial u}\Big|_{u=0+,v}
-\frac{\partial f_{\lambda}}{\partial u}\Big|_{u=0-,v}
-2\cos\alpha s(v)
 \frac{\partial f_{\lambda}}{\partial v}\Big|_{u=0,v}
+\frac{4\big(C_1-C_2\cos\alpha \big)(1-5\xi)}
      {C_2 |v|+C_0}
f_{\lambda}=0.\label{bc1}
\eeq
They are valid on the 4-brane at $u=0$
except at the intersection $u=v=0$.
Similarly,
the boundary conditions
between region $(I)$ and $(IV)$ and between regions $(II)$ and $(III)$ are
\begin{eqnarray}
 \frac{\partial f_{\lambda}}{\partial v}\Big|_{v=0+,u}
-\frac{\partial f_{\lambda}}{\partial v}\Big|_{v=0-,u}
-2\cos\alpha s(u)
\frac{\partial f_{\lambda}}{\partial u}\Big|_{v=0,u}
+\frac{4\big(C_2-C_1\cos\alpha \big)(1-5\xi)}
      {C_1 |u|+C_0}f_{\lambda}=0,\label{bc2}
\end{eqnarray}
\end{widetext}
also, valid on the 4-brane at $v=0$
except at the intersection $u=v=0$.
Each mode must satisfy these conditions with appropriate choices of the integration constants.
In the reminder of this section we will discuss the zero-mode ($m_{\lambda}=0$) solution and address the issue of its localizability on the $3-$brane. The role of the KK excitation will be discussed in a subsequent section.

The zero-mode is described by the homogeneous solutions to Eq.~(\ref{radeqt}) with $m_\lambda=0$
\begin{eqnarray}
f_0= N_0 \left(A^{\alpha_{+}}(u,v)+\gamma A^{\alpha-}(u,v)\right)~,
\end{eqnarray}
with
\begin{eqnarray}
\alpha_{\pm}
=\frac{1}{2}
\left(
-1\pm
\sqrt{1+\frac{4\tilde M^2}{k_1^2+k_2^2}}
\right)\,,
\end{eqnarray}
where $N_0$ is normalization constant. The above solution has to satisfy the condition of normalizability, reality and also the boundary conditions simultaneously.
Normalizability is a restrictive condition and requires $1-\alpha_\pm <0$, selecting the solution with $\alpha_+$. Thus we set $\gamma=0$.
Requiring that the solution is real corresponds to
\beq
\frac{M_s^2}{(k_1^2+k_2^2)}\geq
30\xi -\frac{25}{4}\,,
\label{real}
\eeq
whereas imposing the boundary conditions imply the following relation:
$$
\big(\xi-\frac{1}{5}\big)^2
+{1\over 5}\big(\xi-\frac{1}{5}\big)
-{M_s^2\over 100(k_1^2+k_2^2)}
=0\,.
$$
These two relations imply that $(\xi-1/4)^2\geq 0$, and therefore are satisfied for any value of $\xi$.
However, together with the boundary conditions, normalizability restrict the values of the coupling $\xi$
$$
\xi \leq \frac{1}{10}~, \,\,\, \xi \geq \frac{2}{5}\,.
$$
For $\xi$ varying as above, the normalization constant $N_0$ is
\beq
N_0= \left(\sin\alpha {C_1 C_2\over 2}(1-\alpha_+)(1-2\alpha_+)\right)^{1/2}C_0^{\alpha_+-1}\,.
\eeq
Finally, by looking at the functional form of the solution,
\begin{eqnarray}
f_0= N_0(C_1 |u|+C_2 |v|+C_0)^{-\alpha_{+}}\,,
\end{eqnarray}
it is clear that the zero-mode wave function takes the maximal value at the intersection
and decreases as $|u|$ and $|v|$ are increasing, leading to the localization on the 3-brane, if
$$M_s^2\geq 30\xi(k_1^2+k_2^2)~.$$
It is interesting to note that localization of the zero-mode continue to be valid
also in the minimally coupled case, corresponding to $\xi=0$, despite the fact that
the presence of the intersection does not locally contribute to
the bulk mode function, since the potential $V$ does not contain
any term proportional to $\delta(u)\delta(v)$.

Adding an additional $4-$brane requires some modifications. Here, we wish to argue that the presence of any additional, cut-off $4-$brane in the AdS bulk, does not spoil the above results.
It is clear that the condition of reality (\ref{real}) does not change, but both the boundary and normalization conditions do. Additionally, we have to impose an extra boundary condition at the third $4-$brane.
The boundary conditions at the tensional $4-$branes lead to the following relation
\beq
&&
\left(\alpha_++10(\xi-\xi_c)\right)A^{\alpha_+-1}
\Big|_{u=0}\nonumber\\
&+&\gamma \left(\alpha_-+10(\xi-\xi_c)\right)A^{\alpha_--1}\Big|_{u=0}=0\,.
\eeq
and an analogous relation arising at $v=0$. If we require that the mode is localized, the above relation can be satisfied only by choosing $\gamma=0$ and $\alpha_++10(\xi-\xi_c)=0$, which can be satisfied by tuning the coupling
$$
\xi={1\over 10}\left(1\pm \sqrt{1+\frac{M_s^2}{k_1^2+k_2^2}}\right)\,.
$$
The condition of localization gives, again, the additional constraint
$$
M_s^2\geq 30\xi(k_1^2+k_2^2),
$$
which is always satisfied for the above choice of $\xi$. The final condition that we need to impose are the boundary conditions at the far brane. Since we are treating the brane as tensionless, the boundary conditions are not dictated by the geometry or symmetries, but have to be imposed by hands. One possibility is to impose Robin-type boundary conditions, basically requiring a specific fall-off behavior for the field at the cut-off
\beq
f_0+\sigma_s \partial_X f_0\Big|_{X=\ell}
=0\,,
\label{rob}
\eeq
where $X\equiv C_1u+C_2v$. The coefficient $\sigma_s$ is a constant which is not fixed in the probe brane approximation. Including the brane tension will produce the same type of boundary condition, in general, with
$\sigma_s$ depending on the angles and the other parameters of the model.
By looking at the form of the above boundary conditions, it is clear that the presence of a zero-mode is possible in this case too, but it will require further restrictions on the parameters in the form
$$
(\ell+C_0)+ \sigma_s \alpha_+=0\,,
$$
where nontrivial solutions require $\sigma_s$ to be nonzero.
Another possibility is to place symmetrically at the four sides of the pyramid four cut-off branes and require continuity of the modes at the antipodal branes. For localized zero modes this continuity condition is satisfied.

\subsection{Fermions}
In this section we shall consider the case of higher-dimensional fermions.
In the brane-world the question of fermion localization has received great attention particularly because set-ups with branes and higher-dimensional fermions allow for a geometrical reinterpretations of various features of the standard model like hierarchies of fermion masses or Yukawa couplings with the bonus of a rich phenomenology. A number of articles that studied these issues in the context models with large extra dimensions and warped geometry are Refs.~\cite{Bajc:1999mh,Gherghetta:2000qt,ran,rub,mirab,kogan,neub,ring,sing}.
In the widely studied Randall-Sundrum one and two branes scenario, it was shown that a higher-dimensional fermion has a zero-mode that is naturally
localized on a brane with negative tension. Localization on a positive tension brane is also possible, but at the price of introducing additional interactions, in the form of a mass term of topological origin (basically coupling the fermion with a domain wall).
Here, we wish to reconsider the issue of localization for fermions in the set-up of branes intersecting at angles in anti-de Sitter space, generalizing the previous results for scalars.
We begin by rewriting the fermionic sector of the bulk action (\ref{action})
\beq
S=
\int d^6X \sqrt{-g} \left(
i \bar{\Psi} \underline{\gamma}^M D_M \Psi
+ i \bar{\Psi}M_f \Psi \right)\,,
\label{spact}
\eeq
where we remind that in six dimensions, a Dirac fermion can be represented as an eight component spinor, or equivalently as two four components spinors (in this section we will use the $(z_1,z_2)$ bulk coordinates).
The matrices $\underline{\gamma}_M$ are the Dirac matrices in curved space,
$$
\underline{\gamma}_M = e^{A}_{M}\gamma_A\,,
$$
with $e^{A}_{M}$ being the six-bein, that, in the background spacetime we are considering, can be written as
$$
e^{A}_{M}= A^{-1}\delta^{A}_{M}\,.
$$
In the following we adopt the standard representation for the gamma matrices
\beq
&&\Gamma_{\nu}=
  \left(
  \begin{array}{ccc}
   \gamma_{\nu} & 0  \\
   0  & -\gamma_{\nu}  \\
  \end{array}
  \right),\quad
\Gamma_{z^1}=
  \left(
  \begin{array}{ccc}
   0 & -1  \\
   1  & 0  \\
  \end{array}
  \right),
\nonumber\\
&&\Gamma_{z^2}=
  i\left(
  \begin{array}{ccc}
   0 & 1  \\
   1  & 0  \\
  \end{array}
  \right),
\eeq
where $\gamma_{\nu}$ represent the ordinary four-dimensional gamma matrices.
The six-bein can be expressed as
\beq
e_{\tilde A}{}^B=\left(\delta_{\tilde \mu}{}^{B} \frac{1}{A},\delta_{\tilde z^1}{}^{B} \frac{1}{A},\delta_{\tilde z^2}{}^{B} \frac{1}{A}\right).
\eeq
The non zero component of the spin connection are
\beq
&& \omega_{\mu}{}^{{\tilde z}^1{\tilde \nu}}
=\frac{A_{,z^1}}{A}\delta_{\mu}{}^{\tilde \nu}
,\quad
\omega_{\mu}{}^{{\tilde z}^2{\tilde \nu}}
=\frac{A_{,z^2}}{A}\delta_{\mu}{}^{\tilde \nu},
\nonumber\\
&& \omega_{ z^1}{}^{{\tilde z}^2{\tilde z}^1}
=\frac{A_{,z^2}}{A},\quad
\omega_{z^2}^{{\tilde z}^1{\tilde z}^2}
=\frac{A_{,z^1}}{A}.
\eeq
Finally, the covariant derivatives can be expressed as
\beq
&&D_{z^1} \Psi=\Big(\partial_{z^1}+\frac{A_{,z^2}}{2A} \Gamma_{z^2}\Gamma_{z^1}\Big)\Psi,\nonumber\\
&&
D_{z^2} \Psi=\Big(\partial_{z^2}+\frac{A_{,z^1}}{2A} \Gamma_{z^1}\Gamma_{z^2}\Big)\Psi,\nonumber\\
&&
D_{\mu}\Psi=\partial_{\mu}\Psi +\frac{A_{,z^1}}{2}\Gamma_{z^1}\Gamma_{\mu} +\frac{A_{,z^2}}{2}\Gamma_{z^2}\Gamma_{\mu}\,.
\eeq
The coefficient $M_f$ is an $8\times 8$ matrix, which we write in block diagonal form
$$
M_f= {1\over A}
  \left(
  \begin{array}{ccc}
   M_{11} & M_{12}  \\
   M_{21}  & M_{22}  \\
  \end{array}
  \right)\,.
$$
We assume the terms $M_{11}$ and $M_{22}$ to be constant four-dimensional matrices. As for the off diagonal terms, in order to guarantee that the action respects the $Z_2-$symmetry across the four branes, the form for the four-dimensional matrices $M_{12}$, $M_{21}$ has to be chosen appropriately. Similarly to the Randall-Sundrum case, such terms can be parameterized as follows
\beq
M_{12}
&=&
\kappa_1
\Big(-\frac{A_{,z^1}}{A}
    +\frac{iA_{,z^2}}{A}
\Big),
\nonumber\\
M_{21}
&=&
\kappa_2
\Big(\frac{A_{,z^1}}{A}
    +\frac{iA_{,z^2}}{A}
\Big)\,.
\eeq
Rescaling the spinor field $\Psi$ as
\beq
\Psi
=A^{-5/2}
  \left(
  \begin{array}{cc}
\psi_1     \\
   \psi_2  \\
  \end{array}
  \right)\,,
\eeq
the fermion action (\ref{spact}) can be written as follows
\begin{widetext}
\beq
S=  
i\int d^6X \left(\bar{\psi}_1\gamma^\mu\partial_{\mu}\psi_1 +\bar{\psi}_2\gamma^\mu\partial_{\mu}\psi_2
+\bar{\psi}_1 D^{*}\psi_2+\bar{\psi}_2 D^{}\psi_1
+\bar{\psi}_1 M_{11}\psi_1-\bar{\psi}_2 M_{22}\psi_1
+\bar{\psi}_1 M_{12}\psi_2-\bar{\psi}_2 M_{21}\psi_1 \right)\,,
\eeq
where we introduced the two-dimensional differential operator $D=\partial_{z_1}+i\partial_{z_2}$. At this point it is convenient to decompose the higher-dimensional spinor into its four-dimensional and transverse components
\beq
\Psi&=&
\int d\lambda
\left(
 \begin{array}{ccc}
  f_{1}^{(\lambda)}(z^1,z^2) \psi_{1}^{(\lambda)}(x^{\mu}) \\
  f_{2}^{(\lambda)}(z^1,z^2) \psi_{2}^{(\lambda)}(x^{\mu}) \\
  \end{array}
\right)
\eeq
where $\psi_{i}^{(\lambda)}$ is a four-dimensional Dirac spinor.
Requiring the functions $f_{1}^{(\lambda)}(z^1,z^2)$ and $f_{2}^{(\lambda)}(z^1,z^2)$ to be orthonormal,
\beq
\int dz^1 dz^2
f_{0}^2
=1\,,\,\,\,\,
\int dz^1 dz^2
f_{\lambda}f_{\lambda'}
=\delta(\lambda-\lambda')\,,
\eeq
and to satisfy the following set of equations
\beq
D f_{1}^{(\lambda)}- M_{21}f_{1}^{(\lambda)}&=&im_{2}^{(\lambda)}f_{2}^{(\lambda)}\,,\label{d1}\\
D^* f_{2}^{(\lambda)}+ M_{12}f_{2}^{(\lambda)}&=&im_{1}^{(\lambda)}f_{1}^{(\lambda)}\,,\label{d2}
\eeq
the following expression for the action is easily obtained,
\beq
S=
i\int d^4X d\lambda
\left(
\bar{\psi}_1^{(\lambda)}\gamma^\mu\partial_{\mu}\psi_1^{(\lambda)}
+\bar{\psi}_2^{(\lambda)}\gamma^\mu\partial_{\mu}\psi_2^{(\lambda)}
+im_{1}^{(\lambda)}\bar{\psi}_1^{(\lambda)} \psi_2^{(\lambda)}
+im_{2}^{(\lambda)}\bar{\psi}_2^{(\lambda)} \psi_1^{(\lambda)}
+\bar{\psi}_1^{(\lambda)} M_{11}\psi_1^{(\lambda)}
+\bar{\psi}_2^{(\lambda)} M_{22}\psi_2^{(\lambda)}
\right)\,.
\eeq
\end{widetext}
If we now associate the spinors $\psi_{1}^{(\lambda)}$ and $\psi_{2}^{(\lambda)}$ with the left- and right-handed components of a spinor $\psi^{(a)}$
\beq
\psi_{1}^{(\lambda)}\equiv \psi_{L,a}^{(\lambda)} = {1\over 2}(1-\gamma_5)
\psi_{(a)}^{(\lambda)}\,,\nonumber\\
\psi_{2}^{(\lambda)}\equiv \psi_{R,a}^{(\lambda)} = {1\over 2}(1+\gamma_5)
\psi_{(a)}^{(\lambda)}\,.\nonumber
\eeq
we arrive at the following action for the spinor $\psi^{(a)}$
\beq
S=i\int d^4X d\lambda
\left(
\bar{\psi}_{(\lambda)}^{(a)}\gamma^\mu\partial_{\mu}\psi_{(\lambda)}^{(a)}
+iM^{(a)}_{(\lambda)}\bar{\psi}_{(\lambda)}^{(a)} \psi_{(\lambda)}^{(a)}
\right),
\eeq
with $M^{(a)}_{(\lambda)}=m_{1}^{(\lambda)}+m_{2}^{(\lambda)}$. We have assumed that the four-matrices $M_{11}$ and $M_{22}$ are proportional to the identity matrix with the same proportionality constant.
The masses $M^{(a)}_{(\lambda)}$ are the eigenvalues of (\ref{d1})-(\ref{d2}), that can be found after imposing appropriate boundary conditions.
Above, we have fixed the chirality of $\psi_{1}^{(\lambda)}$ and $\psi_{2}^{(\lambda)}$. This still leaves us the freedom to fix the chirality opposite to the above choice, identifying $\psi_{1}^{(\lambda)}$ and $\psi_{2}^{(\lambda)}$ with the right- and left-handed components of a fermion $\psi_{(\lambda)}^{(b)}$ of different species. This fact can be used to localize separately different chirality modes, generalizing the analogous result for the Randall-Sundrum background.

In the zero-mode case, the masses $M^{(a)}_{(\lambda)}$ are zero, thus the Eqs (\ref{d1})-(\ref{d2}) simplify to
\beq
D f_{1}^{(\lambda)}- \kappa_1
\Big(-\frac{A_{,z^1}}{A}
    +\frac{iA_{,z^2}}{A}
\Big) f_{1}^{(\lambda)}&=&0\,,\label{d11}\\
D^* f_{2}^{(\lambda)}+ \kappa_2
\Big(\frac{A_{,z^1}}{A}
    +\frac{iA_{,z^2}}{A}
\Big)f_{2}^{(\lambda)}&=&0\,.\label{d22}
\eeq
The solution to the above equations can be easily found
\beq
f_{1}^{(\lambda)}&=& A^{\kappa_1}\,,\,\,\,f_{2}^{(\lambda)}= A^{\kappa_2}\,,\nonumber
\label{fs}
\eeq
and the boundary conditions are trivially satisfied on the $4-$branes.
The condition of normalization, instead, requires that $\kappa_1>1$ and
$\kappa_2>1$. In such case, the power-law fall-off of the solutions (\ref{fs}) implies the localizability of the massless zero-modes. Alternatively, we can require that $\kappa_1>1$ and $\kappa_2<1$, thus the left-handed component will be localized, while the other component will not be normalizable. Identifying the species $a$ with neutrinos and the species $b$ with anti-neutrinos, then one can achieve localization at the intersection only of neutrinos and antineutrinos of appropriate chirality.

In the case of an additional $4-$brane, we only have to impose boundary conditions are the far brane. As before we chose general Robin boundary conditions
\beq
f_{1}^{(\lambda)}+\sigma_{1}\partial_x f_{1}^{(\lambda)} \Big|_{x=\ell}&=&0\,,\nonumber\\
f_{2}^{(\lambda)}+\sigma_{2}\partial_x f_{2}^{(\lambda)} \Big|_{x=\ell}&=&0\,.\nonumber
\eeq
As in the scalar of vector case, the above conditions imply some tuning between the various coefficients
\beq
\ell+C_0 - \kappa_1 \sigma_{1} &=&0\,\\
\ell+C_0 - \kappa_2 \sigma_{2} &=&0\,,
\eeq
which one can adjust to achieve localization of definite chiralities. Adding symmetrically four cut-off branes and gluing the modes at the antipodal sides at the bottom of the pyramid will be automatically satisfied by the zero-mode solutions.

In conclusion to this section we mention that if we want to localize a massless fermion zero-mode, the fact that we have intersecting brane is not more advantageous than in the Randall-Sundrum case, as one may expect. Basically, also in this case the (mass) term $M_{\Psi}$ has to be chosen appropriately. It is known, in fact, that massless bulk fermions can be localized on a brane only if
coupled to some domain wall. In this case, there would be a normalizable zero-mode. This zero-mode is topological and its existence does not depend on the detailed profile of the scalar field across the brane  \cite{jacky}.\\


\begin{widetext}

\subsection{Gauge Bosons}

We now turn to the case of gauge bosons.
We begin by briefly recalling the Randall-Sundrum case and refer the reader to Ref.\cite{Gherghetta:2000qt,pomarol} where this case was studied for the two-brane case, and to Ref. \cite{Bajc:1999mh} where the one brane set-up was analyzed.
In the two-brane case, a massless $U(1)$ Gauge field has a massless and constant lowest mode state that in contrary to the graviton or
scalar is not localized on any of the branes.
In the massive case, there is no massless zero-mode and a constant mode still exists, but its mass has become of order of the mass of the bulk field. Also in the one brane set-up there is no localized zero-mode, neither other nontrivial solutions to the equation of motion for the vector field give localized modes.

Let us now consider the case of a massive $U(1)$ gauge boson in the intersecting branes set-up, whose action is the gauge sector of (\ref{action}).
Let us notice that we consider the most general case in which the field is coupled also to the curvature. This corresponds to adding some mass boundary terms that will depend on the coefficients $\chi$ and $\tau$. The ordinary massless, minimally coupled case is obtained by taking the appropriate limit.

The equation of motion for the gauge field $A_M$ is
\beq
{1\over \sqrt{-g}}\partial_M\left(\sqrt{g}g^{MN}g^{PQ}F_{NQ}\right)
+\left(M_A^2-\chi R\right)g^{PQ}A_{Q}-\tau R^{PQ}A_{Q}=0\,.
\eeq
Using the metric (\ref{metric}), we perform the dimensional reduction by decomposing the fields as
\beq
A_{\mu}=
\frac{1}{A}f_0 V^{(0)}_{\mu}
+\int d\lambda
\frac{1}{A}f_{\lambda}V^{(\lambda)}_{\mu}\,.
\eeq
By choosing the gauge $\partial_\mu A^\mu=0,\, A_u=A_v=0$, requiring the radial part of the modes, $f_{\lambda}$ to satisfy
\beq
\Big[
{\partial^2\over \partial u^2}+{\partial^2\over \partial v^2}
-{2 \cos \alpha \over s(u)s(v)}{\partial^2\over \partial u \partial v}
-2\cos\alpha \delta (u)s(v)\partial_v
-2\cos\alpha \delta(v) s(u)\partial_u
-V_A(u,v)
+m_{\lambda}^2\Big]f_{\lambda}=0,
\label{meq}\eeq
and imposing the following normalizability conditions
\beq
\int du dv\sqrt{\gamma} f_{0} f_{0}=1\,,\,\,\,\,\,\int du dv\sqrt{\gamma}
 f_{\lambda} f_{\lambda'}=\delta(\lambda-\lambda')\,,
\eeq
one obtains, a four-dimensional effective field theory in the form of a superposition of a zero-mode plus a tower of massive vector fields
\beq
S=
\frac{1}{2}
\int d^4x
V^{(0)}_{\alpha}
\eta^{\alpha\beta}
\eta^{\mu\nu}\partial_{\mu}\partial_{\nu}
V_{\beta}^{(0)}
+
\frac{1}{2}
\int d\lambda
\int d^4x
V^{(\lambda)}_{\alpha}
\eta^{\alpha\beta}
\Big[
\eta^{\mu\nu}\partial_{\mu}\partial_{\nu}
+
m_{\lambda}^2
\Big]V_{\beta}^{(\lambda)}\,,
\eeq
In the above Eq. (\ref{meq}), we have defined the potential $V_{A}$ as
\beq
V_A(u,v):=
{\tilde M}_A^2 A^2(u,v)
- h_{A1}(u,v) \delta(u)
-  h_{A2}(u,v) \delta(v)~
+8\chi
\Big(\frac{\pi}{2}-\alpha\Big)
\delta(u)\delta(v)
,\label{potential_A}
\eeq
where ${\tilde M}_A^2
:= M_A^2+ (2-30\chi-5\tau\big)(k_1^2+k_2^2)$, and
\beq
h_{A1}(u,v):=
{2\big(C_1-C_2\cos\alpha\big)(1-10\chi-\tau)}A(u,v)\,,\,\,\,\,
h_{A2}(u,v):=
{2\big(C_2-C_1\cos\alpha\big)(1-10\chi-\tau)}A(u,v)\,.
\eeq
Analogously to the scalar case, the boundary conditions can be obtained by integrating across the boundaries.
An easy computation gives, integrating across regions $(I)$ and $(II)$ or $(III)$ and $(IV)$, the following relations
\beq
 \frac{\partial f_{\lambda}}{\partial u}\Big|_{u=0+,v}
-\frac{\partial f_{\lambda}}{\partial u}\Big|_{u=0-,v}
-2\cos\alpha s(v)
 \frac{\partial f_{\lambda}}{\partial v}\Big|_{u=0,v}
+\frac{2(1-10\chi-\tau)\big(C_1-C_2\cos\alpha \big)}
      {C_2 |v|+C_0}
f_{\lambda}\Big|_{u=0,v}=0,\label{bc3}
\eeq
valid on the 4-brane at $u=0$ except on the intersection $u=v=0$.
Similarly,
integrating across regions $(I)$ and $(IV)$, or across regions $(II)$ and $(III)$, one arrives at
\begin{eqnarray}
 \frac{\partial f_{\lambda}}{\partial v}\Big|_{v=0+,u}
-\frac{\partial f_{\lambda}}{\partial v}\Big|_{v=0-,u}
-2\cos\alpha s(u)
\frac{\partial f_{\lambda}}{\partial u}\Big|_{v=0,u}
+\frac{2(1-10\chi-\tau)\big(C_2-C_1\cos\alpha \big)}
      {C_1 |u|+C_0}f_{\lambda}\Big|_{v=0,u}=0,\label{bc4}
\end{eqnarray}
\end{widetext}
valid on the 4-brane at $v=0$ except on the intersection $u=v=0$. Each mode  must be regular and has to satisfy the above boundary conditions along with the normalizability conditions.
In the remainder of the section, we will discuss the existence and localizability of the zero-mode, that, if it exists, will be described by a homogeneous solution to Eq.\,(\ref{meq}), which can be written as
$$
f_0=N_0 \left(A^{\nu_+}(u,v)+\gamma A^{\nu_-}(u,v)\right)\,
$$
where
$$
\nu_\pm = {1\over 2}\left(-1\pm \sqrt{1+{4{\tilde M}_A^2\over k_1^2+k_2^2}}\right)\,.
$$
This case is basically a repetition of the scalar case, so we will be brief. Requiring the reality of the wave function implies
\beq
\frac{M_A^2}{k_1^2+k_2^2} \geq -{9\over 4} +5(6\chi+\tau)\,,
\label{re2}
\eeq
whereas, the condition of normalizability needs $\gamma=0$. Requiring localization for the zero-mode gives $\nu_+>1$ and hence
$$
\frac{M_A^2}{k_1^2+k_2^2}  >30\chi+5\tau\,.
$$
Finally, satisfying the boundary conditions gives a relation between the mass of the gauge field and the couplings
$$
\frac{M_A^2}{k_1^2+k_2^2}=
2\tau +(10\chi+\tau)^2\,.
$$
Combining the above relations gives a condition which is easily satisfied by choosing appropriately the couplings $\chi$ and $\tau$
\beq
10\chi+\tau<0\,,\,\,\,\mbox{or}\,\,10\chi+\tau>3\,.
\label{ssa}
\eeq
Notice that for the inequality to be satisfied at least one of the couplings has to take negative values.
The localizability of the zero-mode on the $3-$brane can be easily checked by looking at the sign of $\nu_+-1$. Positive sign correspond to a power-law fall-off and thus to a localized mode, and this case is realized for the above choice of the couplings (\ref{ssa}). Also in the case  of massless bulk fields, the localization of the zero-mode can be achieved by taking $\tau<0$ and $\chi<-\tau/6$.

If we add an extra brane the bulk solution becomes
\begin{eqnarray}
f_0= N_0 \left( A^{\nu_{+}}(u,v)+\gamma A^{\nu_{-}}(u,v)\right)~.
\end{eqnarray}
The boundary conditions at the $4-$branes lead to the following relation
\beq
&&A^{\nu_+}\left(1-10\chi-\tau-\nu_+\right)\Big|_{u=0}\nonumber\\
&+&
A^{\nu_-}\left(1-10\chi-\tau-\nu_-\right)\Big|_{u=0}=0\,,\nonumber
\eeq
which can be satisfied, consistently with the requirement that the solution is localized, only for $\gamma=0$ and
$$1-10\chi-\tau-\nu_+=0\,,$$
which can be satisfied if
$$
\frac{M_A^2}{k_1^2+k_2^2}=(10\chi+\tau)^2+2\tau\,.
$$
Using the above relation along with the condition of localization, $\nu_+>1$, gives a relation for the couplings
$$
10\chi+\tau<0\,,\,\,\,10\chi+\tau>3\,.
$$
The remaining condition to be imposed is the boundary condition at the far brane. As in the case of scalar fields, the only consistent choice with having a
nonconstant localized zero-mode is that of Robin boundary conditions
\beq
f_0+\sigma_A \partial_X f_0 \Big|_{X=\ell}=0\,,
\label{robA}
\eeq
where the notation is the same as before and $\sigma_A$ is constant. The condition that a mode has to satisfy is then
$$
(\ell+C_0)+ \sigma_A \nu_+=0\,.
$$
As in the scalar case, adding an cut-off brane does not change the localization properties, but restrict the range of parameters for which a localized zero-mode exists.

\section{Discussion}

In this paper, we considered a geometrical set-up which consists of a six-dimensional anti-de Sitter bulk space, cut by two $4-$branes that intersect along a $3-$brane, where all the branes have tension.
This type of brane arrangements uses the properties of both models with large and warped extra dimensions, and thus allows to
take advantage of the virtues of both.
For instance, one can resolve the hierarchy problem by taking infinitely extended AdS bulk.
The effective Planck mass is given by
\beq
M_4^2&=&{2M_6^4\over 3 L^{2}}\sin\alpha~,
\nonumber
\eeq
where
$$
L^2:=\left(k_1\cos\alpha_1+k_2\sin\alpha_1\right)\left(k_1\cos\alpha_2-k_2\sin\alpha_2\right)~.
$$
The above formula tells that $M_4$ is sensitive to the values of the intersecting angles $\alpha_1$ and $\alpha_2$, and the curvature scales $k_1$ and $k_2$. Thus, one can see that by varying the curvature scales, $k_1$ and $k_2$ and the intersecting angles, $\alpha_1$ and $\alpha_2$, it is possible to realize scenarios with gravity in the TeV range. This can be achieved by taking the curvature scale $L$ in the submillimeter range.

In the limiting case of very small intersection angle, $\alpha_1,\alpha_2\sim 0$, one can achieve a smaller Planck scale keeping the curvature scale of $O(1)$.
In Refs.~\cite{origamido,fizran}, it was shown that ordinary four-dimensional gravity can be reproduced on the $3-$brane at the intersection. The role of the correction has also been looked at, and although a rigorous proof is not easy to obtain, Ref.~\cite{origamido} clearly motivates that these are small.
In this work, we have considered a higher-dimensional field theory propagating on the above intersecting brane model, and
discussed under which conditions a localized zero-mode sector exists.
For simplicity, we considered the case of six dimensions, the extension to ten dimension being straightforward.
Our computation generalizes previous results obtained in the case of the Randall-Sundrum one- and two-brane models, and extends the results of Refs.\cite{origamido,fizran}, where the localization has been discussed for the graviton zero-mode.
We have essentially shown that a higher-dimensional version of the standard model can be put in the bulk and, if couplings to the curvature are appropriately chosen (this is equivalent to give to the fields appropriate boundary mass terms due to the presence of the delta functions in the curvature), the zero-mode may be localized. Treating the graviton perturbations as a massless, minimally coupled scalar field, it is possible to see that the conditions given in Sec.~III are satisfied in this case.

We also considered the case when other 4-branes cut the bulk at some distance from the intersection.
As we have seen, in the probe brane approximation, the cut-off brane does not have a significant effect on the field localization on the 3-brane, and this is because the
localization property is determined
by the presence of the potential barrier around
the 3-brane and the boundary conditions
at the intersecting 4-branes, irrespective of the global
topology of the bulk space.

The phenomenology of this class of models is, in principle, interesting. The effective Planck mass can be as low as a TeV scale,
while keeping the bulk infinitely extended, implying the existence of a continuum of KK modes. For this reason, it is important to estimate the size of the coupling of KK modes to brane localized fields. However, this is not straightforward and the reason is simple to understand.
For simplicity, let us take a look at the case of a scalar field
and introduce new bulk coordinates $X := C_1 |u|+C_2 |v|$ and $Y := C_1 |u|-C_2 |v|$.
In this coordinate system,
each KK mode can be written as a superposition of the bulk solutions
\begin{widetext}
\beq
f_{\lambda}
&=&\int_0^{q_{\rm max}}\,\,dq
\sqrt{X+C_0}
\nonumber\\
&\times&
\Big\{
a_{1q}\cos\Big[K(X+C_0)-qY\Big]J_{\nu}
\Big(
Q_{\lambda}(q)\big(X+C_0\big)
\Big)
+a_{2q}
\sin\Big[K(X+C_0)-qY\Big]J_{\nu}
\Big(
Q_{\lambda}(q)\big(X+C_0\big)
\Big)
\nonumber\\
&+&
b_{1q}\cos\Big[K(X+C_0)-qY\Big]Y_{\nu}
\Big(
Q_{\lambda}(q)\big(X+C_0\big)
\Big)
+b_{2q}
\sin\Big[K(X+C_0)-qY\Big]Y_{\nu}
\Big(
Q_{\lambda}(q)\big(X+C_0\big)
\Big)
\Big\}
\label{fqkk}
\eeq
where $J_{\nu}$ ($Y_{\nu}$) is the Bessel (Neumann) function of order $\nu$
and $a_{iq}$, $b_{iq}$ are real coefficients.
The dimensionless constants and parameters are defined by
$\nu:=\sqrt{\tilde M^2/(k_1^2+k_2^2)+ 1/4}$,
$\mu:=m_{\lambda}/\sqrt{k_1^2+k_2^2}$
and
\beq
Q_{\lambda}(q):=\sqrt{\mu^2
      -\frac{4q^2 C_1 C_2\sin^2\alpha}{k_1^2+k_2^2}},\quad
K(q):= \frac{C_1^2-C_2^2}{k_1^2+k_2^2}q.
\label{inf}
\eeq
\end{widetext}
In the above expression, the integrand in (\ref{fqkk}) represents a general solution to the mode equation without boundary terms.
However, when imposing the boundary conditions directly on the integrand, one can show that this form for the modes will not satisfy them, yet being a complete set. Thus, in the same way as Ref.\cite{origamido} does for the graviton zero-mode, we can express a solution as a superposition of the bulk modes, and impose the boundary conditions on the convolutions. This is the meaning of the integral over $q$. Another remark regards the upper bound on the integral (in formula (\ref{fqkk})). In principle the integral should extend up to infinity, however, the requirement of regularity of the mode function, or specifically on the Bessel functions, implies $q$ to be smaller than a maximal value
$q_{\max}
=\sqrt{
m_{\lambda}^2/
(4C_1 C_2 \sin^2\alpha)
}\,$.
The coefficients $a_{iq}$ and $b_{iq}$ ($i=1,2$) are integration constants
which should be determined by the boundary conditions Eq.~(\ref{bc1})
and Eq.~(\ref{bc2}).
The coefficients $a_{iq}$ and $b_{iq}$ are functions of $q$ and the boundary conditions are effectively integral equations in such functions. The above boundary conditions reduce to Eq.(4.22) of Ref.~\cite{origamido}, where the case $M_s=0$, $\xi=0$, $\nu=5/2$, corresponding to the graviton, was discussed. We were not able to find general solutions for the coefficients $a_{iq}$ and $b_{iq}$, but in the same limit as in Ref.~\cite{origamido} (near the intersection and for small mass) and choosing the parameters in such a way that $\nu$ is a half-integer, it is possible to find solutions to lowest order in $q$. In such case the couplings of a localized mode with a KK excitation are suppressed by a factor $m^2L^2/M_4$. In the opposite limit, of masses much larger then the curvature scale, the suppression is by a factor $1/M_4$.

The KK modes satisfy a Schrodinger-like equation, with the potential given by (\ref{potential}).
The reason for such suppression, then, is easily understood by looking at the profile of the potential.
The shape of the potential is shown in Figs. 1-3, where one can see the presence of a barrier around the intersection. Such a barrier protects the $3-$brane from KK contamination, as the low-lying massive excitations are suppressed there, giving rise to suppressed overlap between the KK wave function and the localized modes. More important corrections come from the large mass modes, as the couplings are not suppressed. Once again, the arguments of Ref.~\cite{origamido} can be repeated for some choice of parameters, near the intersection, leading to small correction to the interaction potential.

Adding an additional cut-off brane in the bulk, at a finite distance $\ell$ from the intersection, would have the effect of discretizing the KK spectrum. In this case, upon a specific choice of the parameters (perpendicular branes, and $C_1=C_2$), it is possible to determine the masses for the low-lying part of the spectrum. As one may expect, the masses are inversely proportional to the distance of the cut-off brane from the intersection.
Placing the cut-off brane at a distance comparable with the AdS curvature scale, will give masses in the TeV range. These modes, as in the low-lying part of the continuum case, are innocuous since the overlap with localized zero modes is negligible due to the presence of the potential barrier which protects the brane giving rise to suppressed couplings between the excited and the localized modes.
The above discussion is clearly not complete and to make precise statements, like for example compute accurately the correction to the Newtonian potential or precise bounds on the parameters from particle phenomenology, numerical solutions to the integral equations are required. The arguments of Ref.~\cite{origamido} for the graviton can be applied to the scalar, gauge, and fermion case, thus suggesting that constructing a viable phenomenology is possible.


Stability is also an important issue to consider.
As we have seen,
in all the (scalar, fermion and gauge boson) sectors of perturbations,
no tachyonic mode appears in the spectrum.
Thus, the system discussed in this paper
is expected to be stable against perturbations.


In this paper, we investigated the conditions that would allow to place the
standard model in the bulk of a intersecting brane model in a six-dimensional AdS space. Similarly to the Randall-Sundrum case, in order to localize the zero-mode sector at the intersection, the couplings between the fields and the curvature have to satisfy specific relations. In the case of fermion fields, the mass term has to be chosen appropriately, if one wishes to respect the $Z_2$-symmetries of the geometry. In this case, by tuning the parameters of this mixing matrix, it is possible to localize separately different chiral modes.

\section*{Acknowledgements}
We thank Takahiro Tanaka for discussions.
AF acknowledges the support of JSPS through Grants Nos. 19GS0219, 20740133. The work of MM was supported by the project TRR 33 {\it The Dark Universe} at ASC and by the Korea Science and Engeneering Foundation (KOSEF) grant No. R11-2005-021, founded by the Korean Government (MEST) through the Center for Quantum Spacetime (CQUeST) of Sogang University.
We also acknowledge the Global COE Program ``The Next
Generation of Physics, Spun from Universality and Emergence" from the
Ministry of Education, Culture, Sports, Science and Technology (MEXT) of
Japan.

\end{document}